\documentclass[10pt]{article}
\pdfoutput=1  
\usepackage{graphicx,rotating,hyperref,slashed,amsmath,xcolor,amssymb,amsfonts,colortbl,cite}
\makeatletter
\hypersetup{colorlinks,bookmarksopen,bookmarksnumbered,
linkcolor=blus,pdfstartview=FitH,urlcolor=rossos,citecolor=verde}
\allowdisplaybreaks

\def\lsim{\mathrel{\rlap{\lower3pt\hbox{\hskip0pt$\sim$}}
   \raise1pt\hbox{$<$}}}         
\def\gsim{\mathrel{\rlap{\lower4pt\hbox{\hskip1pt$\sim$}}
   \raise1pt\hbox{$>$}}}         

 \newcommand{\sfootnote}[1]{}
\definecolor{bluc}{cmyk}{1,1,0,0.1}
\definecolor{rossoCP3}{cmyk}{0,.88,.77,.40}
\definecolor{rosso}{cmyk}{0,1,1,0.4}
\definecolor{rossos}{cmyk}{0,1,1,0.55}
\definecolor{rossoc}{cmyk}{0,1,1,0.2}
\definecolor{verdes}{cmyk}{0.92,0,0.59,0.4}

\hypersetup{colorlinks, bookmarksopen, bookmarksnumbered,
citecolor=verdes, linkcolor=bluc, pdfstartview=FitH, urlcolor=rossos}

\newcommand{\mio}[1]{}

\definecolor{Gray}{gray}{0.95}

\newcommand{\bema}{\begin{bmatrix}}
\newcommand{\ema}{\end{bmatrix}}

\usepackage{multicol}
\usepackage{color}
\definecolor{rosso}{cmyk}{0,1,1,0.4}
\definecolor{rossos}{cmyk}{0,1,1,0.55}
\definecolor{rossoc}{cmyk}{0,1,1,0.2}
\definecolor{blu}{cmyk}{1,1,0,0.3}
\definecolor{blus}{cmyk}{1,1,0,0.6}
\definecolor{bluc}{cmyk}{1,1,0,0.1}
\definecolor{verde}{cmyk}{0.92,0,0.59,0.25}
\definecolor{verdec}{cmyk}{0.92,0,0.59,0.15}
\definecolor{verdes}{cmyk}{0.92,0,0.59,0.4}

\def\circa#1{\,\raise.3ex\hbox{$#1$\kern-.75em\lower1ex\hbox{$\sim$}}\,}

\newcommand{\beq}{\begin{equation}}
\newcommand{\eeq}{\end{equation}}

\newcommand{\bea}{\begin{eqnarray}}
\newcommand{\eea}{\end{eqnarray}}
\newcommand{\be}{\begin{equation}}
\newcommand{\ee}{\end{equation}}
\newfam\rsfsfam
\def\mathscr#1{{\fam\rsfsfam\relax#1}}

\def\circa#1{\,\raise.3ex\hbox{$#1$\kern-.75em\lower1ex\hbox{$\sim$}}\,}
\makeatletter

\def\hhref#1{\href{http://arxiv.org/abs/#1}{arXiv:#1}} 

\newcommand{\doi}[1]{\href{http://dx.doi.org/#1}{[doi]}}

\def\hhref#1{\href{http://arxiv.org/abs/#1}{arXiv:#1}} 
 
\def\art{\@ifnextchar[{\eart}{\oart}}
\def\eart[#1]#2#3#4#5#6{{\rm #2}, {\em #3 \bf #4} {\rm (#6) #5} ({\em #1})}

\def\article{\@ifnextchar[{\earticle}{\oarticle}}
\def\oarticle#1#2#3#4#5#6{{\rm #1}, {\em ``#6''}, {\rm #2 #3 (#5) #4}}
\def\earticle[#1]#2#3#4#5#6#7{{\rm #2}, {\em ``#7''}, {\rm #3 #4 (#6) #5}  [\hhref{#1}]}
\def\hepart[#1]#2{{\rm #2, \em#1}}
\def\heparticle[#1]#2#3{#2, {\em ``#3''} [\hhref{#1}]}

\newcounter{alphaequation}[equation]
\def\thealphaequation{\theequation\hbox to
0.6em{\hfil\alph{alphaequation}\hfil}}
\def\eqnsystem#1{
\def\@eqnnum{{\rm (\thealphaequation)}}
\def\@@eqncr{\let\@tempa\relax \ifcase\@eqcnt \def\@tempa{& & &} \or
  \def\@tempa{& &}\or \def\@tempa{&}\fi\@tempa
  \if@eqnsw\@eqnnum\refstepcounter{alphaequation}\fi
\global\@eqnswtrue\global\@eqcnt=0\cr}
\refstepcounter{equation} \let\@currentlabel\theequation \def\@tempb{#1}
\ifx\@tempb\empty\else\label{#1}\fi
\refstepcounter{alphaequation}
\let\@currentlabel\thealphaequation
\global\@eqnswtrue\global\@eqcnt=0 \tabskip\@centering\let\\=\@eqncr
$$\halign to \displaywidth\bgroup \@eqnsel\hskip\@centering
$\displaystyle\tabskip\z@{##}$&\global\@eqcnt\@ne
\hskip2\arraycolsep\hfil${##}$\hfil& \global\@eqcnt\tw@\hskip2\arraycolsep
$\displaystyle\tabskip\z@{##}$\hfil
\tabskip\@centering&\llap{##}\tabskip\z@\cr}
\def\endeqnsystem{\@@eqncr\egroup$$\global\@ignoretrue} \makeatother

\definecolor{fiorentina}{rgb}{.5,0,.5}

\begin{document}

\vspace{1truecm}
 
\begin{center}
\boldmath

{\textbf{\Large An exact solution for  a rotating black hole  in modified gravity}}

\unboldmath

\unboldmath

\bigskip\bigskip

\vspace{0.1truecm}

{\bf Francesco Filippini, Gianmassimo Tasinato}
 \\[8mm]
{\it  Department of Physics, Swansea University, Swansea, SA2 8PP, United Kingdom }\\[1mm]

\vspace{1cm}

\thispagestyle{empty}
{\large\bf\color{blus} Abstract}
\begin{quote}
Exact solutions describing rotating black holes 
can offer   important  tests for alternative   theories of gravity, 
  motivated by  the dark energy and dark matter problems. We present an  analytic rotating black hole
 solution  for  a class of vector-tensor theories of  modified gravity, 
  valid for arbitrary values of the
 rotation parameter.
  The new configuration is  characterised by parametrically large  deviations from the Kerr-Newman geometry, controlled by  non-minimal couplings between vectors and gravity. It 
 has an oblate  horizon in Boyer-Lindquist coordinates, and it can rotate more rapidly and have a larger ergosphere than   black holes  in General Relativity (GR) with the same asymptotic properties. We  analytically investigate the features of the innermost stable circular orbits for massive objects on the equatorial plane, and show that stable orbits lie  further away from the black hole horizon with respect
 to rotating black holes in GR. We also 
  comment on   possible  applications of our findings  for the extraction of rotational energy from the black hole.

\end{quote}
\thispagestyle{empty}
\end{center}

\setcounter{page}{1}
\setcounter{footnote}{0}



\section{Introduction}

The new era of gravitational wave astronomy opens new opportunities for investigating with great precision the 
 physics and dynamics of extreme compact objects, as black holes and neutron stars (see e.g. \cite{Yunes:2016jcc}). It will allow us to study
  for the first time the properties of fundamental  interactions in a strong gravity regime, and test theories
  of gravity that are alternative to Einstein General Relativity (GR) \cite{Clifton:2011jh}. The mysteries of cosmological dark energy and dark matter  motivates 
   attempts to modify GR. 
    For example, it is important to explore the possibility to find
      theories admitting accelerating cosmological solutions with no need of a small positive   cosmological constant (see e.g. \cite{copeland}). The 
    theoretically  most interesting frameworks include  scenarios automatically equipped with screening features, as chameleon \cite{khoury-weltman} or Vainshtein \cite{vain} mechanisms. Screening mechanisms are able to locally hide  the effects of  additional light degrees of freedom besides GR's spin-2 field, and
  reproduce the predictions of Einstein gravity in a weak-field, spherically symmetric regime: see \cite{Joyce:2014kja} for a review. 
  The study of the properties of black hole solutions in these scenarios can  provide new strong gravity tests for these theories, possibly manifesting   sizeable deviations from GR. We focus here on theories with additional degrees of freedom   non-minimally 
  coupled with gravity through
  derivative interactions. Such interactions are essential for an implementation of  Vainshtein screening
   mechanism. In the scalar-tensor case, the prototypes for such set-up are Galileons \cite{Nicolis:2008in} and Horndeski \cite{Horndeski:1974wa}
   theories.  The study of spherically symmetric black hole solutions in these scenarios have lead various interesting results, reviewed e.g. in \cite{Herdeiro:2015waa}. 
   We focus here on vector-tensor versions of these theories, dubbed vector Galileons, or generalized Proca \cite{gripaios,tasinato,heisenberg}. Various examples of static, spherically symmetric black hole configurations have been found, and the study of compact objects as neutron stars have been recently  developed \cite{vecblr1,vecblr2}. 
   
   In this work, we present and study examples of  rotating black hole solutions with regular
   horizons for  vector Galileons. Rotating black hole configurations which deviate from the Kerr family are hard to obtain in 
   any theory of gravity, and only  few examples of exact solutions are known in modified gravity frameworks. Solutions
   are known for scalar-tensor theories \cite{Babichev:2014tfa,Moffat:2015kva},  also  with  a complex scalar \cite{Herdeiro:2014goa}, 
   and in the context of Einstein dilaton Gauss-Bonnet
   theories \cite{Pani:2011gy,Kleihaus:2015aje}.  Slowly rotating solutions
   in Horndeski theories are discussed in \cite{Maselli:2015yva}.  
   See  \cite{Berti:2015itd}
    for a comprehensive review, and \cite{Johannsen:2011dh,Konoplya:2016jvv} for  useful parameterisations of  possible 
    deviations from the Kerr family of
    black holes in the context of arbitrary theories of gravity. 
    Yet, given the fact that most astrophysical black holes are spinning, it is important to
    pursue the effort to  determine and analyze 
    explicit 
    rotating
   black hole configurations in theories alternative to General Relativity.  An additional theoretical reason to study
    rotating configurations is the fact that these objects break spherical symmetry (being at most axially symmetric). Hence
    they are  an ideal set-up to start investigating screening mechanisms -- as for example the Vainsthein mechanism -- that
    are well studied and are known to be  efficient only   for spherically symmetric systems.

    We determine exact solutions describing rotating black holes, by applying a disformal transformation
      on a version of the  Kerr-Newman (KN) solution of the Einstein-Maxwell theory of gravity. The resulting
      configuration solves
       the equations of motion associated to a particular vector Galileon action, and 
        is parameterically  different from a   KN system. 
         Having exact solutions allows us to analytically investigate distinctive  properties of 
          spinning black holes in our theory.   Our 
         configurations are characterized by three asymptotic charges, the black hole mass, angular momentum, and
         vector charge. 
        We show that the black hole horizon
        is oblate in Boyer-Lindquist coordinates, since its radial 
        position depends on the polar angle. This is a feature that can make our black hole
           distinguishable from KN solutions, whose horizon  lies
        at constant value of the radius in such coordinate system.  The black hole maximal spin can be parametrically
        larger than KN configurations, for the same values of the asymptotic charges. 
         The solution admits also a `massless' limit of black hole with zero mass, but with a vector charge which ensures
         the existence of a regular horizon. 
         The study of equatorial circular trajectories admits an analytical treatment. We show that probe
       massive objects can  rotate faster than in the KN family of solutions.  Innermost stable circular orbits
        (ISCOs) lie  further away from the black hole horizon with respect
 to rotating black holes in GR. We also 
  comment on   possible  applications of our findings  for the extraction of rotational energy from the black hole.

\section{Set-up}

We build a modification of Einstein gravity which includes  additional  vector degrees of freedom,
belonging to the class of theories dubbed vector-tensor Galileons \cite{gripaios,tasinato,heisenberg}. Such
vector degrees of freedom can be 
associated with dark forces  motivated by dark matter or dark energy model building.  Our  aim  will be 
to investigate
 new rotating
black hole solutions for the theory we consider. 

In order to construct a modified gravity action,
 our  starting point is  a standard  Einstein-Maxwell system, described by an
action
\beq \label{eimact}
S_{EM}=\int{d^4x\sqrt{-\tilde{g}}\left[ \frac{\tilde{R}}{4}-\frac{1}{4}\tilde{F}^{\mu\nu}\tilde{F}_{\mu\nu}\right]}\,.
\\
\eeq
Although we call the previous action an `Einstein-Maxwell' system, as we are going to discuss the vector fields appearing in eq \eqref{eimact} should not be identified with standard electromagnetism, but with additional dark vector forces.  
We use
a mostly plus   metric signature, and Weinberg's conventions for the  Riemann and derived tensors. Our results are expressed in natural units, setting $c=1$, $\hbar=1$, and $4 \pi G =1$. The theory enjoys an 
Abelian gauge invariance, $A_\mu\,\to\,
 A_\mu+\partial_\mu \xi$ for any arbitrary function $\xi$. It propagates four degrees of freedom, two in the transverse traceless
  tensor sector, and two in the  transverse vector sector. 
 Exact black hole solutions for  the equations of motion associated  for this theory are well studied, and include  static Reissner-Nordstr\"om and rotating
Kerr-Newman (KN)  configurations. They satisfy no-hair theorems, which state that black holes are uniquely defined
in terms of their mass,  angular momentum, and charge.

 To generate new solutions,
we act on this action with a disformal transformation \cite{Bekenstein:1992pj,Bettoni:2013diz,Kimura:2016rzw} involving vector fields and parameterized by a 
real constant
$\beta$, which  plays a key role in what follows:
\bea \label{def-dist}
\tilde{g}_{\mu\nu} (x) &=&g_{\mu\nu}  (x) -\beta^2 A_{\mu} (x)  A_{\nu} (x) \, ,
\\
\tilde{A}_\mu  (x) &=& A_\mu  (x) +\partial_\mu \alpha  (x)  \label{def-distA} \, ,
\eea
for any arbitrary function $\alpha(x)$ (the transformation of the vector simply reflects the gauge freedom of the original 
theory, and $\alpha$ does not appear in the final formulae). 
More general disformal/conformal transformations can be considered (see for example \cite{Kimura:2016rzw}), but transformation \eqref{def-dist} is sufficient
for our purposes. 
The disformed metric is invertible, with inverse given by
\be
\tilde{g}^{\mu\nu}=g^{\mu\nu}+\beta^2 \gamma_0^2 A^{\mu} A^{\nu}\,\,\,\,\,\,\,\,\,\,\text{with } 
\,\,\,\,\,\,\,\,\gamma_0^2=\frac{1}{1-\beta^2 A^{\mu}A_{\mu}} \, .
\ee
Up to total derivatives, the disformed action reads
\beq
S_{disf}=S_{EH}+S_{matt}=\int{d^4x \sqrt{-g}(\mathcal{L}_{EH}+\mathcal{L}_{matt})}\\ \, ,
\eeq
with  Lagrangian densities
\beq
\mathcal{L}_{EH}=\frac{1}{4 \gamma_0}\left[ {R} -\frac{\beta^2}{4}  \gamma_0^2\left( S_{\mu\nu}S^{\mu\nu}-S^2 \right)+\frac{\beta^2}{4}F_{\mu\nu}F^{\mu\nu}+\frac{\beta^4}{2} \gamma_0^2F_{\mu\rho}F_{\nu}^{\,\rho}A^{\mu}A^{\nu} \right]\\
\eeq
and
\beq
\mathcal{L}_{matt}=-\frac{1}{4 \gamma_0}[F^{\mu\nu}F_{\mu\nu}+2 \beta^2 \gamma_0^2F_{\mu\rho}F_{\nu}^{\,\rho}A^{\mu}A^{\nu}],
\eeq
with
\beq
\begin{aligned}
&F_{\mu\nu}= \nabla _{\mu}A_{\nu}-\nabla_{\nu}A_{\mu} \, , \\
&S_{\mu\nu}= \nabla _{\mu}A_{\nu}+\nabla_{\nu}A_{\mu} \hskip0.5cm ,\hskip0.5cm
S\,=\,S_{\mu\nu}g^{\mu\nu} \, .
\end{aligned}
\eeq
In a more compact notation:
\beq \label{disf-act}
S_{disf}=\int{d^4x  \sqrt{-g} \frac{1}{4 \gamma_0}\left[   {R} - \frac{\beta^2}{4}  \gamma_0^2\left( S_{\mu\nu}S^{\mu\nu}-S^2 \right)-\frac{4-\beta^2}{4}F_{\mu\nu}F^{\mu\nu}+\frac{\beta^4 - 4\beta^2}{2}\gamma_0^2F_{\mu\rho}F_{\nu}^{\,\rho}A^{\mu}A^{\nu} \right]} \, .
\eeq
The  system after the disformal transformation describes a vector-tensor theory of gravity which deviates
from the usual Einstein-Maxwell case of eq  \eqref{eimact}  by quantities depending on the disformal 
parameter $\beta^2$.  
Action  \eqref{disf-act} will be our  modified gravity theory of reference  in this work. It contains non-minimal
couplings of the vector to gravity, and derivative self-interactions of the form that usually characterize Horndeski
systems.  The  theory is free of Ostrogradsky instabilities. 
In fact, this system 
  belongs to the class of theories dubbed vector-tensor Galileons or generalized Proca, which have
been investigated at length  for their distinctive properties for cosmology \cite{cosmoapp}, field theory \cite{fieldapp}, and black holes \cite{vecblr1,vecblr2}.  Disformal transformations are known to preserve the causality properties
of the  theory one starts with \cite{Bekenstein:1992pj}: hence since the Einstein-Maxwell theory is causally well behaved, one does not expect  instabilities or causal pathologies in the theory described by eq \eqref{disf-act}.

 Although the two actions \eqref{eimact} and \eqref{disf-act} are related
  by a disformal transformation, they are not  equivalent when  additional matter, minimally coupled with gravity, is included into  the system. Hence, interpreting the system of \eqref{disf-act} as part of a more general action including matter fields, we can expect that its physical consequences can be  different from an Einstein-Maxwell set-up. We will elaborate
  on these points in the next Sections.
  
  %
Action   \eqref{disf-act}  breaks an Abelian gauge invariance,  since the Lagrangian depends explicitly on the gauge potential.   On the other hand, 
it inherits some memory of the 
   original  gauge symmetry of action \eqref{eimact}. In fact, the final action is invariant, up to total derivatives,
 under the more general gauge transformation
 \bea \label{trn1}
 g_{\mu\nu}&\to& g_{\mu\nu}+\beta^2 
 \,
  \partial_\mu \chi \,A_\nu+\beta^2 
  \,
  \partial_\nu \chi \,A_\mu+\beta^2
 \partial_\mu \chi  \partial_\nu \chi \, ,
 \\  \label{trn2}
 A_\mu  &\to&A_\mu  +\partial_\mu \chi \, ,
 \eea
with $\chi$ an arbitrary scalar function. It is easy to check  that the    quantities
$\tilde g_{\mu\nu}$ and $\tilde A_\mu$,
 eqs \eqref{def-dist} and  \eqref{def-distA}, are invariant under the simultaneous transformations \eqref{trn1}, \eqref{trn2}. Hence   the final action,
which can be expressed  in terms of 
the combinations $\tilde g_{\mu\nu}$ and $\tilde A_\mu$,
 is invariant under this symmetry. Notice that  the limit $\beta\to0$  reduces, as expected, to standard Abelian gauge symmetry.   Such symmetry ensures that the final action propagates
  four dynamical modes. The absence of a fifth, scalar degree of freedom is a welcoming
 feature for phenomenology, since it can automatically avoid stringent constraints on the existence of long range
 scalar fifth forces. In this work, we focus on determining and analysing 
  new regular rotating black holes for the theory  \eqref{disf-act},  equipped with the new gauge symmetry \eqref{trn1}, \eqref{trn2}, obtained
  by a disformal transformation of the Einstein-Maxwell action. 
 
 
  \section{New rotating solutions in vector-tensor theories of gravity}
 
 In this Section we  show that  a  disformal transformation acting on an appropriate solution of the original
 field equations leads to a new regular, rotating black hole for action \eqref{disf-act}, with a non-trivial
 profile for the vector field turned on.  One might expect 
  that 
  the   non-linearity of field equations, and the fact that we renounce to spherical
  symmetry,  imply that 
  rotating
  configurations in this theory are different  from GR  solutions. We confirm this expectation, showing
  that
   rotating configurations
  with non-trivial vector  profiles have specific properties that make them  distinguishable from
  their GR counterparts. This fact can lead to qualitatively new ways to test modified gravity models, by investigating the  
  properties of their black hole solutions.
  
  
  \smallskip
  
We start from a Kerr-Newman solution of the Einstein-Maxwell action \eqref{eimact}, which can  be expressed
 in Boyer-Lindquist coordinates as 
 \bea
 d s^2&=&
 \left( \frac{dr^2}{\Delta}+d\theta^2 \right) \rho^2 -\left( dt- a \sin^2\theta \,d\phi \right)^2\frac{\Delta}{\rho^2}+\left[ (r^2+a^2)\,d\phi-a\,dt \right]^2\frac{\sin^2\theta}{\rho^2} \, ,\label{vems10}
 \\
 A_{\mu}&=&\left\{-\frac{Q \,r}{\rho^2},A_r(r),0,\frac{a \, Q\, r\,\sin^2\theta}{\rho^2}\right\} \, , \label{vems1}
 \eea
 with:
\beq
\begin{aligned}
&\Delta= a^2 +r^2 -2Mr+Q^2 \, ,\\
&\rho^2= r^2+a^2 \cos^2\theta \, .\\
\label{defDrho}
\end{aligned}
\eeq
 This configuration describes a rotating  charged black hole. See e.g. \cite{Teukolsky:2014vca,Adamo:2014baa} for comprehensive
 reviews on rotating black hole solutions in GR. 
 The constants $M$, $Q$, and $a$ are associated with 
  the black hole mass, charge, and angular momentum. 
  We turn on an arbitrary radial component for the vector potential, which we denote with $A_r(r)$:
 since the original Einstein-Maxwell action respects an Abelian gauge symmetry, $A_\mu\to A_\mu+\partial_\mu \chi$, the radial profile of  $A_r(r)$
 does not affect the geometry, and can in principle be `gauged away'. On the other hand, it plays an important 
 role for our purposes, since the disformed transformation  mixes metric
 and vector degrees of freedom, and does not enjoy the standard Abelian symmetry.
 
 We  apply the disformal transformation of eq \eqref{def-dist} to 
 eqs \eqref{vems10}, \eqref{vems1}. The resulting configuration is solution
 of the vector-tensor theory  \eqref{disf-act}, describing a rotating
  system. 
   Potential  problems  arise though,
since   the resulting geometry  generically has naked singularities or other pathologies 
  not covered by horizons. Alternatively it is
   not  asymptotically flat, or it results 
  too complicated to be  of any use. 
    On the other hand, the radial vector profile $A_r(r)$ in eq \eqref{vems1} influences the
   geometry after the disformal transformation, and we can use this fact at our advantage.  
 We determined a specific profile for the  radial vector component,
 which leads to a regular, asymptotically flat  black hole configuration:
 \bea
 A_r(r)\,= \,\frac{ Q r}{\Delta(r)} \, .
  \eea
  It would be interesting to understand whether other radial vector profiles give regular solutions. 
 With this choice for the radial vector component, 
 the disformed metric contains off-diagonal components %
 %
 %
 $dr\,dt$ and $dr\,d \phi$ components
and the final metric reads 
\beq
ds^2=g_{tt}\,dt^2+g_{rr}\,dr^2+g_{\theta\theta}\,d\theta^2+g_{\phi\phi}\,d\phi^2
+2\,g_{tr}\,dt\,dr+2\,g_{t\phi}\,dt\,d\phi+2\,g_{r\phi}\,dr\,d\phi,
\eeq
with
\beq
\begin{aligned}
g_{tt}\,=\,&-1-\frac{\left(Q^2-2 M r\right) \left( r^2+a^2 \cos^2\theta\right)-\beta^2  Q^2 r^2}{\left( r^2+a^2 \cos^2\theta\right)^2} \\
g_{tr}\,=\,&\frac{-\beta^2\,  Q^2 r^2}{\left(a^2+r^2-2 M r+Q^2\right) \left( r^2+a^2 \cos^2\theta\right)}\\
g_{t\phi}\,=\,&a \sin^2\theta \frac{\left(Q^2-2 M r\right) \left( r^2+a^2 \cos^2\theta\right)-\beta^2  Q^2 r^2}{( r^2+a^2 \cos^2\theta )^2}\\
g_{rr}\,=\,&\frac{\left(a^2+r^2-2 M r+Q^2\right) (r^2+a^2\,\cos\theta^2)+ \beta^2  Q^2 r^2}{ (a^2 +r^2 -2Mr+Q^2)^2}\\
g_{r\phi}\,=\,& \frac{\beta^2\,a\,   Q^2 r^2 \sin^2\theta}{\left(a^2+r^2-2
   M r+Q^2\right) \left( r^2+a^2 \cos^2\theta\right)}\\
g_{\theta\theta}\,=\,& r^2+a^2 \cos^2\theta\\
g_{\phi\phi}\,=\,& \sin^2\theta \left[r^2+a^2-a^2 \sin^2\theta\frac{  \left(Q^2-2 M r\right) (r^2+a^2\,\cos^2\theta)-\beta^2\,Q^2r^2  }{\left( r^2+a^2 \cos^2\theta\right)^2}\right].
\end{aligned}
\eeq
In a more compact notation we can write
\small{
\beq
\begin{aligned}
ds^2=& \,-1-\frac{\left(Q^2-2 M r\right) \rho^2-\beta^2  Q^2 r^2}{\rho^4}\,dt^2-\frac{\beta^2\,Q^2r^2}{\Delta\rho^2}dt\,dr+a\sin^2\theta\frac{(Q^2-2Mr)\rho^2-\beta^2\,Q^2r^2}{\rho^4}dt\,d\phi+\frac{\Delta\,\rho+\beta^2\,Q^2r^2}{\Delta^2}dr^2\\
&\,+\frac{\beta^2\,a\sin^2\theta Q^2r^2}{\Delta \rho^2}dr\,d\phi+\rho^2\,d\theta^2+\sin^2\theta\left[ r^2+a^2-a^2\,\sin^2\theta\frac{(Q^2-2Mr)\rho^2-\beta^2\,Q^2r^2}{\rho^4} \right]d\phi^2
\label{dis-sol1}
\end{aligned}
\eeq}
with $\Delta$, $\rho$ given in eqs \eqref{defDrho}, while the gauge field components read 
\bea
A_\mu&=&
\left(A_t\,,\,A_r\,,\,A_{\theta}
\,,\,A_{\phi}
\right)
\nonumber
\\
&=&\left(
-\frac{Q \,r}{\rho^2}\,,\,\frac{Q\,r}{\Delta}\,,\,0\,,\,
\frac{ Q \,r\,a\, \sin^2\theta}{\rho^2} 
\right) \, .
 \label{eq:gaugevec}
\eea
 This geometry describes a rotating black hole with regular horizon, charged under the vector field:  
 \begin{itemize}
 \item
 the dimensionless quantity $\beta^2$ parametrically controls deviations from
  the Kerr-Newman geometry in eq \eqref{dis-sol1}. 
The geometry depends on three
 integration constants, $M$, $a$, and  $Q$, as well as by the parameter $\beta^2$ associated
  the non-minimal couplings of vector to gravity in action \eqref{disf-act}.  Besides mass and spin, the geometry
  is charged under the vector degrees of freedom that control our modification of gravity (and should not be identified
  with electromagnetism). 
  As we shall learn in the next Sections,  
  the contributions depending
  on $\beta$ modify the structure of the black hole horizons, and has consequences for the properties
  of geodesics of massive particles;
  \item the vector field  profile \eqref{eq:gaugevec} has three  physical 
  components turned on, against the two of the Kerr-Newman configuration. The vector radial component is physical
 in this case, and can not be gauged away without simultaneously changing the geometrical properties
 of the system. 
   \end{itemize} 
 Using the Mathematica 
  package xAct \cite{xAct}, we explicitly checked that eqs \eqref{dis-sol1}  \eqref{eq:gaugevec}  are  a solution of all the equations of motion associated
  with the disformed vector-tensor action.
  It is important to emphasise that although this configuration is disformally related to Kerr-Newman, it is a new solution for the vector-tensor modified gravity theory we are considering.
    The systems \eqref{eimact} and \eqref{disf-act} can have distinct physical
 implications when additional matter is included, minimally coupled with gravity. For example, we will show in Section 
 \ref{sec-orb} that the features of    time-like geodesics in this black hole background are different with respect to rotating black
 holes in GR.   
       The expression for the Ricci scalar  is 
 \beq
R\,=\, 2\,\beta^2\,a^2\,Q^2 \, \cos^2\theta \frac{   \left(a^2  \cos^2 \theta-3 r^2\right)}{\left(r^2+a^2  \cos^2\theta\right)^4} \, .
\eeq
Since the Ricci scalar is non-vanishing,  this configuration is different from the original KN solution, where this quantity is equal to zero.   In Appendix \ref{app-A} we show that asymptotic mass, charge and angular momentum for 
this configuration are $M$, $Q$, and $a/M$. 
The unique geometrical singularity associated with our new
disformed solution \eqref{dis-sol1} is the Kerr 
 singularity at the locus $r^2+a^2  \cos^2\theta=0$. We also checked that, besides this singularity,
  the remaining curvature invariants, obtained contracting Ricci
and Riemann tensors, 
 are everywhere regular. The solution is asymptotically flat since the metric components approach
 flat space at asymptotic infinity, and the curvature invariants asymptotically tend to zero.  In principle we can still use the gauge freedom of eq \eqref{trn1}, \eqref{trn2} to change the profile of vector components, 
  for example turning off the radial quantity $A_r$. However, this gauge operation changes the metric as well, and can lead to singular geometries. For this reason we work with metric and gauge field as expressed in
  eqs \eqref{dis-sol1}, \eqref{eq:gaugevec}.  
  Before discussing in some detail the non-linear features of this disformed configuration, it is interesting to analyse
  the limit of small rotation: at first order in an expansion on the rotation parameter $a$, the  geometry reads
   
  \beq
  \begin{aligned}
  ds^2=&\left(-1+\frac{2 M}{r} -\frac{(1-\beta ^2) Q^2}{r^2}\right)dt^2+\frac{r^4}{\Delta^2} \left(1-\frac{2
   M}{r}+\frac{Q^2 (1+\beta^2)}{r^2}\right)dr^2+r^2 d\theta^2+r^2 \sin^2\theta d\phi^2\\
  & -\frac{2\, \beta^2\,Q^2
   }{\Delta} \,dt\,dr+2\,a\,\sin^2\theta\left[-\frac{2 M}{r}+\frac{ Q^2 (1-\beta^2)}{r^2}\right]\,dt\,d\phi+\frac{2\,\beta^2 \,a\, Q^2   \sin ^2(\theta
   )}{\Delta} dr\,d\phi,
   \end{aligned}
  \eeq
  where $\Delta$ is calculated with $a^2=0$.\\
  In the limit $a \rightarrow 0$, the metric becomes
  \beq
  ds^2= \left(-1+\frac{2 M}{r}-\frac{(1-\beta^2 ) Q^2}{r^2}\right)\,dt^2+\frac{r^4}{\Delta^2} \left(1-\frac{2
   M}{r}+\frac{Q^2 (1+\beta^2)}{r^2}\right)\,dr^2+r^2\,d\theta^2+ r^2 \sin^2\theta\,d\phi^2-\frac{2 \beta^2\,Q^2
   }{\Delta}\,dt\,dr
  \eeq
 which can be considered the disformal Reissner-Nordstrom solution.
 
  It would be interesting to  investigate whether the regime $\beta^2 > 1$ leads to problems for the system (e.g. instabilities 
   of vector degrees of freedom around our geometry). But we do not touch these topics in this work, and explore the properties of the geometry for any value of 
   $|\beta|$. 
 In the next section  we will perform an analytic study of  the structure  of 
 horizons, which have features which make them distinguishable
 from their GR counterparts.
 
  \section{Structure and properties of  horizons}\label{sec-hor}
 
 We  analyse  the  structure of the horizons, where  departures from the Kerr geometry are
  manifest.
The vector Lagrangian  \eqref{disf-act} is non-minimally coupled with gravity, and the non-linear derivative interactions
induce  qualitative deviations from GR rotating solutions. For example, we will learn that our black holes can rotate faster than their
GR counterparts, for the same mass and charge. 

Given a spherical coordinate system $\{t,r,\theta,\phi \}$, the {\it event horizon} corresponds
 to points where the   hypersurfaces of constant $r$ become null, namely
\beq
g^{\mu\nu}\partial_{\mu}r\partial_{\nu}r=g^{rr}=0 \, ,
\eeq
where $\partial_{\mu}r$ is the $1$-form normal to   constant $r$ hypersurfaces.
 A Killing horizon, instead, is the null hypersurface where the length of a Killing vector vanishes,
 $k^{\mu}k_{\mu}=0$. For stationary geometries as our configuration, the
  {\it ergosphere}  corresponds to the Killing horizon of the time translation Killing vector $k^\mu=\{1,0,0,0 \}$.
  A priori, ergosphere and event horizon are  distinct hypersurfaces.
   See
  \cite{Hartle:2003yu} for details.  
  For the standard KN black hole (obtained from solution \eqref{dis-sol1} setting $\beta^2=0$), the exterior event horizon and the ergosphere are situated at
  \beq
\begin{aligned}
r_{hor}^{KN} \,=\,M+\sqrt{M^2-a^2-Q^2} \, ,\\
r_{erg}^{KN}\,=\,M+\sqrt{M^2-a^2\cos^2\theta-Q^2} \, .
\end{aligned}
\eeq
Using Boyer-Lindquist coordinates,  the event horizon is spherical in the Kerr geometry,
 being located to a radial position independent on the angular coordinates.  (Although
the  intrinsic horizon geometry is actually squashed, as manifest using appropriate
 coordinate systems \cite{Hartle:2003yu}).
The ergosphere is instead an ellipsoid. 
 
 In the vector-tensor theory of gravity we are examining,  the  event
  horizons and 
   the ergospheres are hypersurfaces of revolution around the azymuthal coordinate. Their  positions is  
  given by the real positive solutions of   equations

  \bea
&&{\text {\bf Horizon:}}\,\,\,\,\,\,\hskip1cm g^{rr}=0 \,\,\,\, \Rightarrow \,\,\,\,f_h\,\equiv\, \left(r^2+a^2 \cos^2 \theta\right) \left[
r^2-2 M r+ a^2+Q^2
\right]-\beta^2  Q^2 r^2=0  \, ,\nonumber 
\\ \label{evhorp}
\\
&&{\text {\bf Ergosphere:}}\,\,\,\,\,\, 
k^{\mu}k_{\mu}=0 \,\,\,\, \Rightarrow \,\,\,\,
f_e\,\equiv\, 
\left(r^2+a^2 \cos^2\theta\right) \left[
r^2-2 M r+ a^2 \cos^2 \theta+Q^2
\right]
-\beta^2 Q^2 r^2=0 \, .
\nonumber 
\\
\label{ergp}
\eea
\noindent
Equations $f_{h,\,e}\,=\,0$ are  algebraic equations of fourth order in the  coordinate $r$: they can  be solved analytically,  but their
solutions are complicated.
 Depending on whether the sign of the discriminant, they can have four, two or no real roots. 
 We denote  their maximal real roots,   with $r_h$ and $r_e$
respectively: $r_h$ corresponds to the position of the external 
 horizon of the black hole.
 Such expressions  depend on the polar angle $\theta$: in our configuration, neither the ergosphere  nor the horizons 
have spherical shape in Boyer-Lindquist  coordinates.  This is a major qualitative difference
with respect to Kerr-Newman black holes, which have spherical horizons  in Boyer-Lindquist  coordinates.  
  We find that $r_e\ge r_h$:  substituting the value $r_e$ in the expression for
 $f_h$ \eqref{evhorp}, we find $f_h\ge 0$, hence the position of the ergosphere is outside the  horizon.
 The  horizon and ergosphere hypersurfaces touch at the  poles, $\theta\,=\,0,\,\pi$, since
 at these points $f_{h,\,e}$ coincide. 
   The only value of the polar angle where an expression for $r_{h,\,e}$ can be
  analytically obtained is the equator, where we get 
 
  \beq
\begin{aligned}
r_{h}(\theta\,=\,\pi/2)\,= &M+ \sqrt{M^2-a^2-Q^2(1- \beta^2 ) } \, , \\ 
r_{e}(\theta\,=\,\pi/2)\,= &M+ \sqrt{M^2-Q^2(1- \beta^2 ) }.
\end{aligned}
\eeq
 It is simple to show that the external horizon of Kerr-Newman  black hole is always in the interior of the horizon 
 of our disformal black hole configuration, for the same values of mass, charge and angular momentum. Substituting 
  the value $r_h^{KN}$ for  the position of the Kerr-Newman  horizon in the expression \eqref{evhorp} for 
  $f_h$, we find that this quantity is negative, hence it lies inside the disformal black hole horizon.  
   Fig. \ref{figure1}   represents the shape of black hole horizons and ergosphere for our systems, using Boyer-Lindquist
   coordinates.
  
  
   \begin{center}
\includegraphics[scale=0.52]{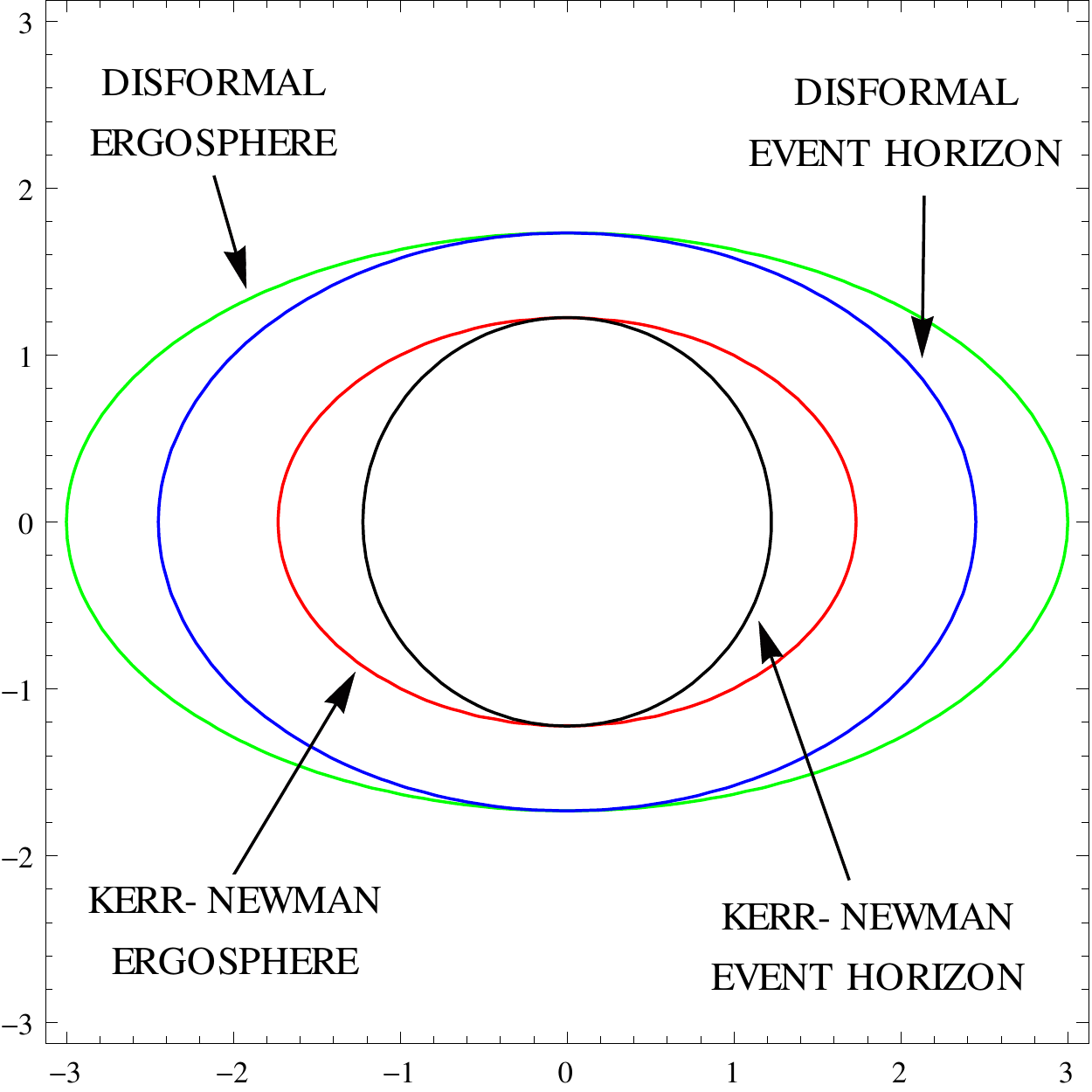}
\begin{figure}[h]
\caption{\it{Pictorial representation of the shape  of horizon and ergosphere for the disformed black hole using Boyer-Lindquist coordinates,  as discussed in the main text. 
}}
 \label{figure1}
\end{figure}
\end{center}

At this point it is interesting looking at the event horizon's angular velocity.\\
Let's consider a photon emitted along the $\phi$ direction at radial distance $r$; so, there are no components of the velocity along the $\theta$ and $r$ directions, and the metric simply reads
\beq
ds^2=g_{tt}\,dt^2+2\,g_{t\phi}\,dt\,d\phi+g_{\phi\phi}\,d\phi^2=0,
\eeq
which can be solved to obtain
\beq
\frac{d\phi}{dt}=-\frac{g_{t\phi}}{g_{\phi\phi}}\pm \sqrt{\left(\frac{g_{t\phi}}{g_{\phi\phi}} \right)^2-\frac{g_{tt}}{g_{\phi\phi}}}.
\eeq   
 Evaluating this expression on the event horizon
 \beq
 f_h=0,
 \eeq
 the term in the square root term vanishes and it simply remains
 \beq
 \frac{d\phi}{dt} |_{r_h}=\Omega_h=\frac{a}{a^2+r_h^2},
 \eeq
  which is the angular velocity of the event horizon. Notice that since the position of the horizon depends
   on the angular coordinate $\theta$, being it solution of eq \eqref{evhorp}, the angular velocity is not constant 
   on all the horizon surface. 
   %

  \subsection{Maximal black hole spin and horizon oblateness}

The modifications of the geometry proportional to the parameter $\beta$ can deform
the horizon, making it oblate also in Boyer-Lindquist coordinates. 
Interestingly, they can also allow
 for  ultra spinning black holes, i.e. black hole configurations that rotate faster than Kerr. 
 It is not easy to analytically study these properties of the geometry, a part from two special cases that we 
 are going to investigate in this subsection.

  \subsubsection*{Case 1: The limit of large value of $\beta^2$}

The size of the disformal horizon increases
with the polar angular coordinate: its smallest value is at the poles $\theta\,=\,0,\,\pi$, while its maximal value
is at the equator $\theta=\pi/2$. The existence of a regular horizon depends on the value of the black hole
spin parameter $a$: beyond an extremal value $a_{max}$, an horizon ceases to exist. 
Such value of $a_{max}$
can be found by studying 
the equation which gives the radial position of the horizon at the poles in Boyer-Lindquist coordinates ($\theta\,=\,0,\,\,\pi$).
We focus on the poles 
 because at this position we get 
the most stringent condition on $a_{max}$.
 The radial horizon equation at the poles
 can be expressed
as 
\be \label{eqhp2}
r^4-2\,M\,r^3+\left( 2 a^2+q\,M^2\right)\,r^2-2 a^2\, M\,r+a^4+\frac{a^2\,M^2\,q}{1-\beta^2} 
\,=\,0 \, ,
\ee
where we define the combination
\be \label{defsq}
q\,\equiv\,\frac{Q^2}{M^2}\left(1-\beta^2\right) \, ,
\ee
which can be positive or negative, depending on  the size of $\beta$.  Analytic
solutions for 
equation \eqref{eqhp2} are cumbersome: but a great simplification occurs in the limit $|\beta|\gg1$. We consider
 this regime
maintaining a  fixed value for the parameter $q$ defined in eq \eqref{defsq}. This implies that we simultaneously take a limit in which black hole charge
$Q$ becomes smaller and smaller, so to maintain $q$ finite. This limit physically correspond to a regime of `strong non-minimal coupling' between vector and gravity in the  action \eqref{disf-act}. 

In this regime, the last term in eq  \eqref{eqhp2}  can be neglected, and the equation admits a simple expression for its
real roots: the external horizon sits at the position
\be
r_{h}^{pol}\,=\,\frac{M}{2} \left(
1+\sqrt{1-q}
+\sqrt{2-q+2\sqrt{1-q}-4\frac{a^2}{M^2}}
\right) \, .
\ee
Requiring a positive argument for the last square root in the previous expression imposes
an upper bound on the rotation parameter $a$: its largest allowed value is
\bea \label{defama}
a_{max}
&=&
\frac{M}{2}
\sqrt{2-q+2\sqrt{1-q}}
\\
&=&\frac{M}{2}
\sqrt{2+\frac{Q^2}{M^2}\left(\beta^2-1\right)+2\sqrt{1+\frac{Q^2}{M^2}\left(\beta^2-1\right)}} \, \, , \label{defama2}
\eea 
where in the second line we use the definition \eqref{defsq}. When 
 $\beta^2\,>\,1$, then $a_{max} \ge M$ (with equality saturated for $Q=0$). 
Recall that $M$ is the maximal value of the spin $a$ for a Kerr black hole (while for Kerr-Newman the maximal spin is
$\sqrt{M^2-Q^2}\,<\,M$). Hence when $\beta^2$ is large,
    our disformed rotating black hole can have  angular momentum
 parametrically larger than in Einstein gravity.
  (Ultraspinning black holes are however possible in theories with more than $3+1$ dimensions \cite{Myers:1986un} {or in modified gravity theories including a complex scalar \cite{Herdeiro:2014goa} or EdGB \cite{Kleihaus:2015aje}}.) 

We obtained this  result in the extremal limit of very large $|\beta|$: it is possible to check numerically that for smaller values of $|\beta|$, 
the value of $a_{max}$ reduces its size with respect to \eqref{defama2}, by a quantity that is proportional to $1/|\beta|$. In any case, we will  use $a_{max}$ of eq \eqref{defama2} 
as reference for our discussion.  We also checked that the conditions on the parameter $a$ for having a 
regular horizon are most stringent at the poles, and less restrictive at other angular positions. In other words, the horizon at the equator could in principle
rotate faster than $a_{max}$ of eq \eqref{defama}, but the requirement of having a regular horizon everywhere does not allow for this.

When the spin parameter attains the value $a_{max}$, the ratio between the radial size of the horizon at the poles
versus the size of the horizon at the equator quantifies the oblateness $\omega$ of the black hole in Boyer-Lindquist
coordinates
\be \label{def=ob}
\omega\,=\,1-\frac{r_h^{pol}}{r_h^{eq}}=\,1-\frac{1+\sqrt{1-q}}{2+\sqrt{2-3 q-2\sqrt{1-q}}}  \,  .
\ee
In the limit of large, negative values of $q$, oblateness approaches the extremal value $\omega_{max}\,
=\,1-1/\sqrt{3}$:  hence in this limit
the radial size of the horizon at the poles 
  is $1/\sqrt 3 \simeq 0.57$ times smaller than the size of the  horizon at the equator, as measured in Boyer-Lindquist coordinates.
  
  \smallskip
  
  We can interpret the horizon properties  we determined  as due to our non minimal  vector-tensor interactions, which are able to contrast strong centrifugal
  forces, and maintain a regular horizon even for large spins, at the price of deforming the horizon shape.

    \subsubsection*{Case 2:  Massless black holes}

Interestingly, if $\beta^2>1$,  our black hole configuration \eqref{dis-sol1} admits  a regular horizon even in the massless 
limit $M\to0$.  In this limit, the solution has two horizons: the radial position of the external
one depends on the polar angle and is given by the expression (valid if $\beta^2>1$)

\be
r_h^2\,=\,\frac12\left[ 
Q^2\left(\beta^2-1\right)-a^2 \left(1+\cos^2\theta\right)+\sqrt{
\left[Q^2 \left(\beta^2-1\right)-a^2 \left(1+\cos^2\theta\right)\right]^2-4 a^2 \left(a^2+Q^2\right)\,\cos^2\theta} 
\right] \, .
\ee
The radial size of the black hole is maximal at the equator ($\theta=\pi/2$) and minimal at the poles ($\theta\,=\,0,\pi$). 
The condition of having a regular horizon at the pole imposes an upper bound 
 on the black hole spin parameter, given by
\be
a_{max}^2\,=\,{Q^2}\,\frac{\left(\beta^2-1\right)^2}{4 \beta^2} \, .
\ee
 It is also simple to compute the black hole oblateness for extremal values of the black hole spin, as done in Case
 1 around eq \eqref{def=ob}. In this case we obtain
 \be
 \omega\,=\,1-\sqrt{\frac{1+\beta^2}{1+3 \beta^2}} \, \,  .
 \ee
Also here the maximal oblateness is $1-1/\sqrt{3}$, showing that for large $\beta$
the radial size of the horizon at the poles 
  is $1/\sqrt 3 \simeq 0.57$ times smaller than the size of the  horizon at the equator. 
  
\section{Equatorial orbits}\label{sec-orb}

The dynamics of  massive and massless fields orbiting  around rotating black holes is a broad 
subject with several ramifications and applications to  astrophysics and cosmology, and it is the first step towards the study  of black hole accretion disks, 
or of
the 
 extraction of rotational energy
from spinning black holes. 
 See \cite{Teukolsky:2014vca} for an enlightening review, and \cite{Bardeen:1972fi} for the original paper studying this family of orbits
 in Kerr configurations. 
 We focus our attention to circular orbits for massive particles in the equatorial plane, examining features
that are peculiar of our disformed rotating black hole.  We make the hypothesis that the particles are only minimally coupled with gravity.  
 Since the geometry is axially symmetric,  stable orbits exist which remain confined on the equatorial  plane. 
 %
 Having an exact form for the metric allows us to point out
distinctive properties of equatorial orbits by simple, analytical considerations, which are a natural generalization
of arguments developed for the Kerr(/Newman) geometry \cite{Bardeen:1972fi}. 

%
%
%
%
%
%
%
The disformal metric (\ref{dis-sol1}) on the equatorial plane ($\theta={\pi}/{2}$) reads

\beq
ds^2=g_{tt}\,dt^2+g_{rr}\,dr^2+g_{\theta\theta}\,d\theta^2+g_{\phi\phi}\,d\phi^2
+2\,g_{tr}\,dt\,dr+2\,g_{t\phi}\,dt\,d\phi+2\,g_{r\phi}\,dr\,d\phi. \label{eq:disfeq}
\eeq

\beq
\begin{aligned}
g_{tt}=&\,-\frac{r (r-2 M)+(1-\beta^2) Q^2}{r^2}\\
g_{rr}=&\,\frac{r^2 \left[a^2+r (r-2 M)+(1+\beta^2 )
   Q^2\right]}{\Delta^2}\\
g_{\theta\theta}=&\,r^2\\
g_{\phi\phi}=&\,r^2+\frac{a^2 \left[r (2 M+r)-(1-\beta^2) Q^2\right]}{r^2}\\
g_{tr}=&\,-\frac{\beta^2 \, Q^2}{\Delta}\\
g_{t\phi}=&\,\frac{a \left[(1-\beta^2 ) Q^2-2 M r\right]}{r^2}\\
g_{r\phi}=&\,\frac{ \beta^2  \,a\,Q^2}{\Delta}
\end{aligned}
\eeq

Such metric is independent from $t$ and $\phi$, hence 
 we can define the \emph{conserved energy per unit mass} $e$ and the \emph{conserved angular momentum per unit mass} $\ell$ along the symmetry axis:

\beq
\begin{aligned}
&e \equiv -k^{\mu}\,u^{\nu}\,g_{\mu\nu} \, ,\\
&\ell \equiv r^{\mu}\,u^{\nu}\,g_{\mu\nu} \, ,
\end{aligned}
\eeq
where $u^{\mu}$ is the $4$-velocity vector, and $k^{\mu}$ and $r^{\mu}$ are the Killing vectors defined in Appendix   \ref{app-A}. 
Using the metric (\ref{eq:disfeq}), the previous equations can be expressed in the following way:
\beq \label{eq:cond1}
\begin{aligned}
&e \,=\, -(g_{tt}\,u^t+g_{t\phi}\,u^{\phi}) \, ,\\ 
&\ell\,=\,g_{\phi t}\,u^t+g_{\phi\phi}\,u^{\phi} \, ,
\end{aligned}
\eeq
with ($\tau$ being proper time)
\beq \label{def4v}
u^{\mu}=\{u^t,\,u^r,\,u^{\theta},\,u^{\phi} \} = \left\{\frac{dt}{d\tau},\,\frac{dr}{d\tau},\, \frac{d\theta}{d\tau},\, \frac{d\phi}{d\tau} \right\}.
\eeq
Inverting the previous relations, one obtains the angular velocity at fixed radial distance from the equator,
\be
\frac{u^{\phi}}{u^t}=\Omega\,=\,\frac{\ell \left[r (r-2 M)+(1-\beta^2 ) Q^2\right]-a \,e
   \left[(1-\beta^2) Q^2-2 M r\right]}{e\, r^4+a^2 \,e \left[r (2
   M+r)-(1-\beta^2) Q^2\right]+a\, \ell \left[(1-\beta^2 ) Q^2-2 M
   r\right]} \, . \label{genOH}
\eeq

\smallskip

Time-like geodesics  associated with massive particles satisfy the
condition
\beq
u^{\mu}\,u^{\nu}\,g_{\mu\nu}=-1 \label{eq:cond2}
\,.
\eeq
To compute the radial position of stable circular time-like trajectories on the equatorial plane, we assume
 that $u^{\theta}=0$, and we can combine equations (\ref{eq:cond1}) and (\ref{eq:cond2}) to obtain the following expression for derivatives of the radial position of a massive particle along proper time
\beq
\frac{r^4}{\Delta^2}\left[ \Delta^2-\beta^4\,Q^4 \right]\left(\frac{dr}{d\tau}\right)^2-\frac{2\,\beta^2\,r^2\,Q^2}{\Delta}\left( a^2 e-a\ell+er^2 \right)\left(\frac{dr}{d\tau}\right)-\left( a^2 e-a\ell+er^2 \right)^2+\left( \Delta-\beta^2\,Q^2 \right)\left( 1+\frac{(a\,e-\ell)^2}{r^2} \right)=0.
\eeq
The previous equation can be recast as
\beq
 \frac{\left[ \Delta^2-\beta^4\,Q^4 \right]}{2\Delta^2}\left(\frac{dr}{d\tau}\right)^2-\frac{\,\beta^2\,Q^2}{r^2\Delta}\left( a^2 e-a\ell+er^2 \right)\left(\frac{dr}{d\tau}\right)+V_{eff(r,e,\ell)}=\frac{e^2-1}{2}  \label{radeq1}
 \eeq
 where we have define an effective potential
\beq
V_{eff}=-\frac{M}{r}+\frac{a^2 \left(1-e^2\right)+\ell^2+(1-\beta^2 ) Q^2}{2 r^2}-\frac{M
   (\ell-a e)^2}{r^3}+\frac{(1-\beta^2 ) Q^2 (\ell-a e)^2}{2
   r^4}.  \label{effpotex}
\eeq
The effective potential above
  is both energy and angular momentum dependent.  Equations
   \eqref{radeq1}, \eqref{effpotex} are what we need to analyse corotating ($\ell>0$) and counterrotating ($\ell<0$)  circular trajectories, associated with objects moving
   in the same or opposite sense of the black hole.

  Before continuing, it is important to notice that the expression for the potential 
\eqref{effpotex} has a structure identical to the KN case, which is recovered sending $\beta^2\to0$. New opportunities arise for our case with non-linear
vector-tensor interactions.  
    In the regime $\beta^2\,>\,1$, the relative sign among different contributions to the effective
    potential \eqref{effpotex} changes with respect to standard KN black holes (in that case, this regime would correspond to an unphysical
    negative square charge).
          Hence this regime is interesting since it can 
 lead to qualitatively new features for circular orbits. 
  As a concrete example, a straightforward computation starting from eq \eqref{genOH} leads to the following
  expression for the angular velocity of massive particles on circular orbits
  
  \be \label{spOH}
\Omega\,=\,\frac{M}{ a \,M\pm\frac{r^2 \,M}{\sqrt{r M+ Q^2 
\,\left({\beta^2-1}\right)
}}} \, ,
\ee
 with $\pm$ indicating orbits corotating/counterrotating with the black hole. If $\beta^2>1$, corotating orbits
 spin faster than the corresponding Kerr-Newman case, for identical values of the asymptotic
 charges $M$, $Q$, $a$.

\smallskip

We now proceed  examining the existence and properties of
 marginally stable innermost circular orbits for test particles moving on the equatorial plane of our geometry, lying
  on a fixed radial position   $\bar r$ in Boyer-Lindquist coordinates.  Such trajectories
are called ISCO \cite{Hartle:2003yu}. 
 To move on a circular orbit of radius  $\bar r$, both the initial radial velocity and the radial acceleration must vanish, hence

\beq
\begin{aligned}
&V_{eff}(\bar r,e, \ell)=\frac{e^2-1}{2} \,\label{circon1} \, ,\\
&\frac{\partial V_{eff} (r,e, \ell)}{\partial r} \Big{|}_{r= \bar r}\, =\,0 \, .
\end{aligned}
\eeq
Furthermore, the condition of stability implies that the particle must be at a minimum point of the potential, namely
\beq
\frac{\partial^2 V_{eff} (r,e, \ell)}{\partial r^2} \Big{|}_{r=\bar r} \geq 0\,, \label{circon3}
\eeq
where equality holds for the marginally stable circular  trajectories that we wish to investigate.
   The previous three  conditions lead to three equations which determine the three quantities $e$, $\ell$, $\bar r$ characterizing 
    marginally stable innermost trajectories.    
In order to express formulae in the simplest possible way, it 
  is convenient to rescale our quantities
  as follows
  
  \be
 x\,=\,
 \frac{\bar r}{M}
 \hskip0.5cm
 ,
  \hskip0.5cm
  \hat a \,=\,
  \frac{a}{M}
 \hskip0.5cm
 ,
  \hskip0.5cm
   q\,=\,\frac{ Q^2 }{M^2}\,\left({1-\beta^2}\right) \, .
 \ee
  Notice that $q$ can be negative, if $\beta^2\,>\,1$: a negative  $q$ 
   can not be obtained in a Kerr-Newman-like configuration, and is distinctive of our disformed black hole geometry. 
  We  can re-express and solve the 
  three equations \eqref{circon1}-\eqref{circon3} (with inequality saturated) in terms of the three unknown quantities
  $e$, $\ell$, $x$.  After various algebraic manipulations, we get the following expressions
  for  $e$ and $\ell$ in terms of $x$:
\bea
e^2&=&\frac{q (3-4 x)+x (3 x-2)}{x (3 x-4 q)} \, ,
\\
\frac{\ell^2}{M^2}&=&\frac{ \hat a^2 (3 q-2 x)+x \left(4 q^2-9 q x+6 x^2\right)}{x (3
  x-4 q)} \, .
\eea
Only the angular momentum of the particle depends on the black hole spin $\hat a$: there are two possibilities to consider, positive or negative $\ell$ -- corresponding to trajectories corotating or counterrotating with respect to the black hole.   
 The equation determining the radial position $x\,=\,\bar r/M$ of the ISCO is 
 an algebraic equation of sixth degree
 \bea
&x^6-12  x^5- 6  \left( \hat a^2-3
   q-6\right)x^4 -4  \left(-2 \hat a^2 q+7 \hat a^2+2 q^2+27 q\right)x^3 + 3 \left(3 \hat a^4+10 \hat a^2 q+43 q^2\right)x^2 \\
 &  - 24  \left( \hat a^4
   q+4 \hat a^2 q^2+3 q^3\right)x+16 q^2 ( \hat a^2+q )=0 \, .
    \eea \label{sixteq}
 
 There are  up to six real solutions to the previous equation, depending on the values of $a$, $q$:  fully
   analytical expressions are in general  unavailable. We are interested to examine 
    the case $q\le0$, distinctive of our black hole configuration.
     We demand that the ISCOs
     lie outside the position of the external black hole horizon, located at
    $$
 x\ge x_{hor}\,=\,1+\sqrt{1-\hat a^2-q} \, .
 $$

  We   study corotating and counterrotating configurations for representative choices of $ q$
   for the rest of this section, starting from a brief review of ISCOs for Kerr black holes ($q=0$). 
   
   \begin{itemize}
   
   \item[$\blacktriangleright$] {\bf Case $q=0$: the Kerr black hole.} 
The case $q=0$ reduces the system to a Kerr black hole. Equation  \eqref{sixteq} simplifies considerably, and leads  to 
two branches of physical solutions, corresponding to a corotating and a counterrotating orbit: see the
textbook \cite{Hartle:2003yu} for an excellent account.  We plot
 in Figure  \ref{figureA1} the radial position of the orbit $x\,=\,\bar r/M$ versus black hole spin $a/M$:
 corotating orbits become closer and closer to the horizon as we increase the spin of the black hole (see black curve
 on the left panel), while counterrotating orbits become more and more distant. Corotating orbits can touch the horizon for the extremal
 value  of the spinning parameter $a$. 
The  binding energy of the object 
 per unit rest mass -- given by the quantity  $(1-e)$  --
versus black hole spin have opposite behaviour: the binding energies
of corotating objects on ISCOs increase as the black hole spin increases, while the same quantity decreases for counterrotating
objects. 
The maximal binding energy for an object in a  corotating orbit on an extremal
 black hole, $a=M$ reaches the value $1-e\,=\,0.42$.

\begin{center}
\includegraphics[scale=0.84]{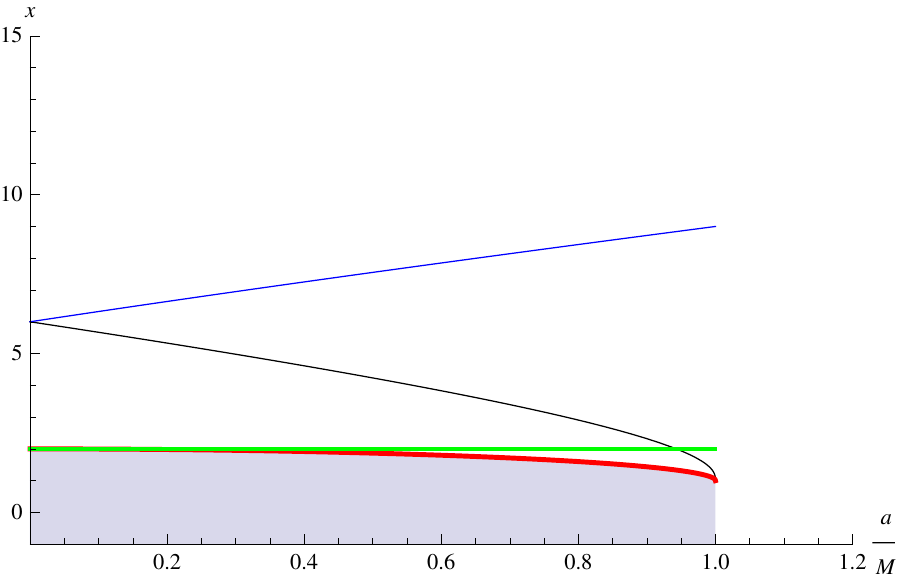}
\includegraphics[scale=0.84]{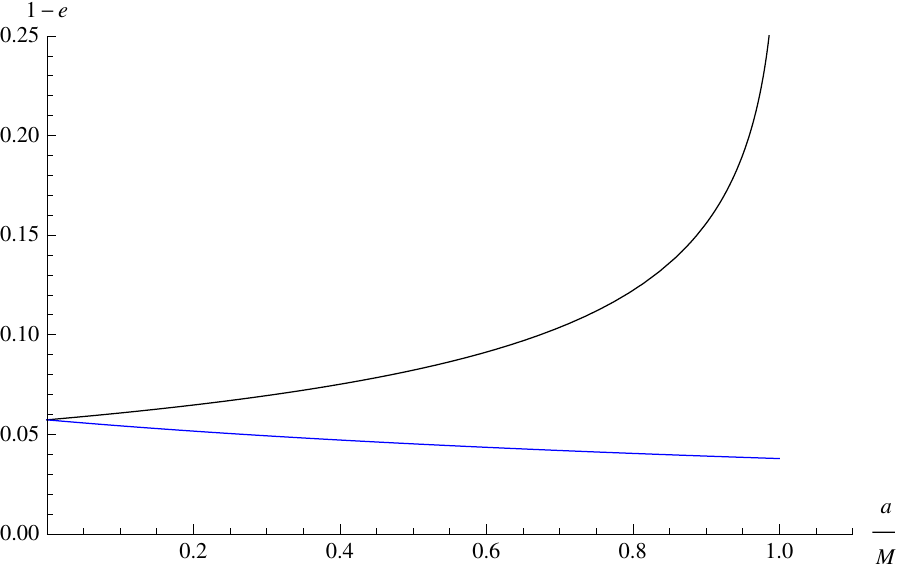}\end{center}
\begin{figure}[h]
\caption{\it{ Panel on the left: 
Innermost
 circular orbits for a Kerr black hole: radial position of the orbit $x\,=\,\bar r/M$
versus black hole spin $a/M$.  There are two branches of solutions, corresponding to corotating (black) and counterrotating  (blue) ISCOs.  The red line corresponds to the horizon, and the shaded
part the forbidden region inside the black hole horizon. The green line is the boundary of the ergosphere.  Panel on the right:  binding energy $1-e$ versus black hole spin $a/M$.
}}\label{figureA1}
\end{figure}

   \item[$\blacktriangleright$] {\bf Case $q=-0.2$}:
 When $q$ is small and negative, the properties of ISCO 
 trajectories are qualitatively  similar to the case of Kerr.  An important difference is that corotating stable orbits can neverand touch the horizon,
 not even for extremal values of the spinning parameter, which is
  $a_{max}\,=\,1.05\,M$ for $q=-0.2$. The binding energy for corotating ISCO never exceed values of order 20 per cent in this case. Notice that, as discussed
 in Section \ref{sec-hor},
  the equatorial region of the black hole {\it could} rotate faster than $a_{max}$, and still be well defined: we represent with dashed red curve the additional interval of black hole spin that would be allowed at the equator. On the other hand,  the requirement of having a regular horizon everywhere, including at the poles, restricts  the value of $a$ to be smaller than
  $a_{max}$. 
 \begin{center}
\includegraphics[scale=0.84]{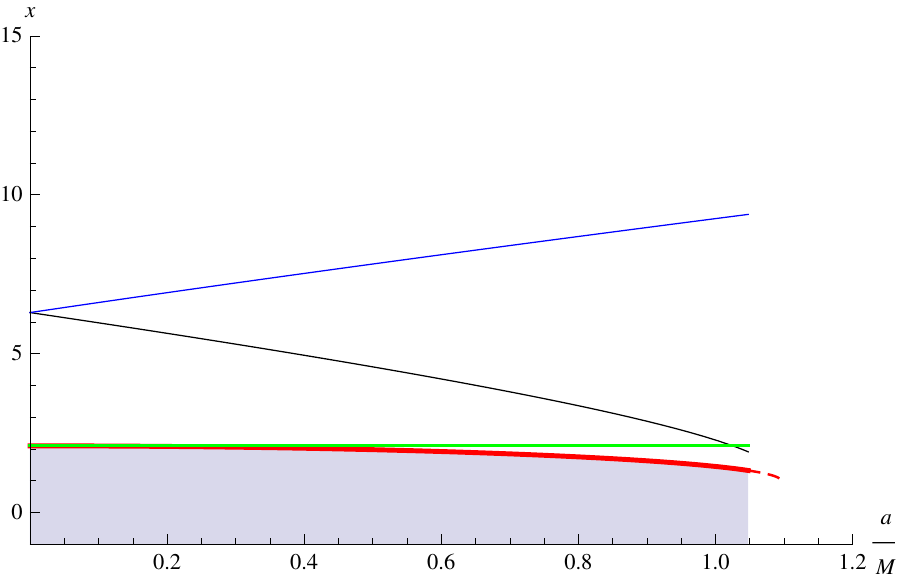}
\includegraphics[scale=0.84]{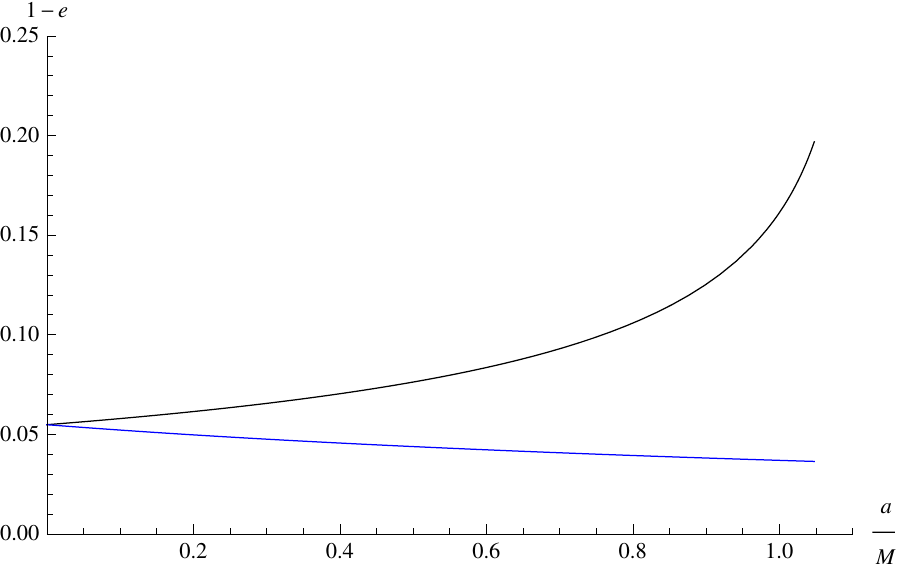}
\begin{figure}[h]
\caption{\it{
Innermost  circular orbits (panel on the left) and  binding energy versus black hole spin (panel on the right)
 for a  black hole  with $q=-0.2$. Color codes as in Figure \ref{figureA1}.
 The corotating ISCO does not touch the black hole horizon, 
    even for extremal values of the black hole spin. 
 }}\label{figureA2}
\end{figure}
\end{center}
   \item[$\blacktriangleright$] {\bf Case $q=-5$}: 
 When $q$ becomes more negative, the features we  noticed for small $q$ become more accentuated.  
For $q=-5$, the maximal value of the rotation parameter is $a_{max}\,=\,1.72 M$. Corotating ISCOs are  far from the black hole horizon, even for extremal 
values of $a_{max}$.
The maximal value of the binding energy on a corotating ISCO is of few per cent.
\begin{center}
\includegraphics[scale=0.84]{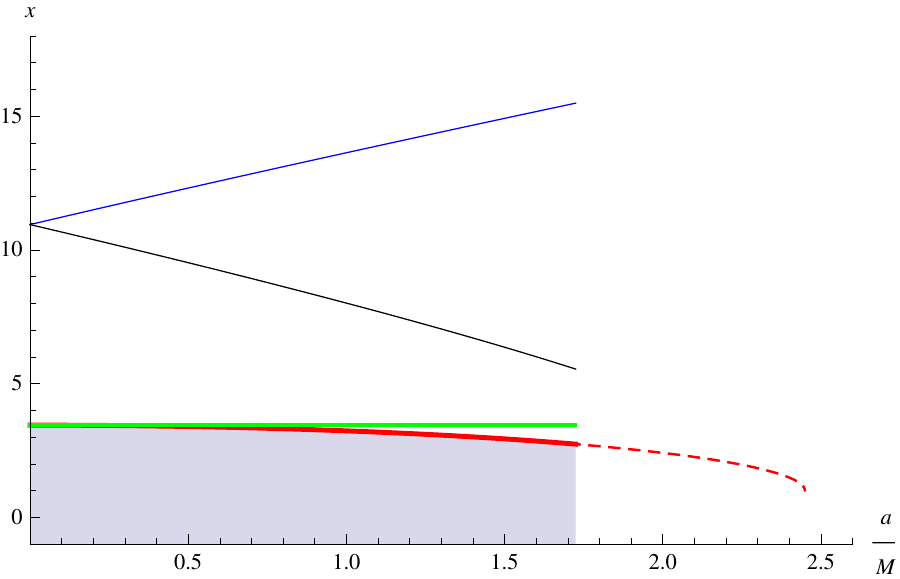}
\includegraphics[scale=0.84]{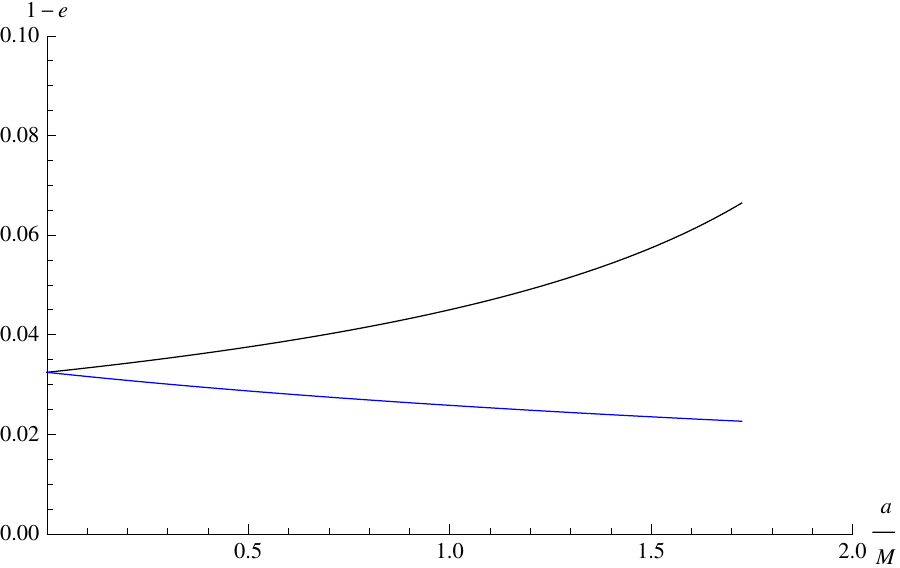}
\begin{figure}[h]
\caption{\it{ISCOs (left) and   binding energies  (right) versus black hole spin, 
 for a  black hole  with $q=-5$. Color codes as in Figure \ref{figureA1}.
 }}\label{figureA3}
\end{figure}
\end{center}

\end{itemize}


\bigskip

\bigskip

\newpage
Our findings might be relevant for  studying extraction of rotational energy from the 
disformed black hole -- a broad topic that 
  goes outside the scope of the present paper, but that we can start to qualitatively
   touch here. For large values of $|\beta|$, the size of the ergoregion can be  greater  than for Kerr black holes: at the equator, the radial size of the ergosphere is 
   \be
   r_{erg}\,=\,M\left(1+\sqrt{1+\frac{Q^2}{M^2} \left(\beta^2-1\right)}\right) \, ,
   \ee
   hence it can be well larger than $2\,M$ for  $|\beta| \gg 1$. 
Massive particles in the ergosphere   can have negative energy \cite{Hartle:2003yu}. 
This implies that a Penrose process \cite{Penrose1,Penrose2}
 can be in principle devised: a massive object  -- e.g. a star -- can break
 into two fragments through tidal effects   within the ergosphere. One part -- with negative energy -- falls into the black hole; the other escapes, carrying
 away more energy than the initial object, and slowing down the black hole rotation. Since the ergosphere region can be large in our set-up, it might be easier
 to extract energy through this process~\footnote{Although it might be not too efficient since only  
  unstable circular orbits are contained in the ergoregion, and the falling objects might not find the correct orbital configurations to make the mechanism feasible.}. 
  Other  mechanisms 
   for energy extraction can be 
   applied in our context, as black hole
    superradiance \cite{zeld} or 
some version of the Blandford-Znajek mechanism \cite{Blandford:1977ds}, possibly using our vector interactions. 
In fact, it is known that the efficiency of energy extraction can be increased for charged black holes \cite{bhat}.
Moreover, as we have learned around equation \eqref{spOH}, for
our configurations the angular velocity  of massive particles
on circular orbits can be parametrically larger than in Kerr, possibly making more efficient
 the mechanism behind the idea of black hole colliders \cite{Banados:2009pr}.
We plan to  return to discuss these topics in 
 a separate publication.
 
\section{Discussion}

In this work we determined and analysed exact solutions for rotating black holes in a specific example of vector Galileons, a theory
of modified gravity motivated by the dark energy problem, and  that involves additional vector degrees of freedom besides a spin-2 graviton. The set-up have many features in common with several modified gravity models, including derivative
self-couplings and non-minimal couplings with gravity. 
 We
determined our new black hole solution applying an appropriate  disformal transformation to a system related to the  Kerr-Newman solution
of  Einstein-Maxwell gravity, and discussed various physical implications that differentiate this system from known
rotating solutions in General Relativity. 

The rotating black holes configurations we determined are valid also for large  values of the black hole spin parameter. Their
deviation from a Kerr-Newman solution is parametrically controlled by a dimensionless quantity  associated
with  non-minimal couplings between vectors and gravity. The black holes are characterised by three
asymptotic conserved quantities -- black hole mass, spin, and vector charge. The black hole horizon
is oblate in Boyer-Lindquist coordinates, since its radial position depends on the polar coordinates: this makes
a difference with standard Kerr-Newman solutions, for which the horizon is at fixed radius in Boyer-Lindquist coordinates.
 We shown that the maximal value for the black hole spin can be parametrically larger than in the Kerr-Newman
 family of solutions, for the same value of asymptotic charges.  We noticed that our solutions maintain a regular horizon even in the limit of zero mass for the black hole. 
 We then studied  
equatorial circular trajectories around our solutions, which admit an analytical treatment. We shown that probe
       massive objects can  rotate faster than around  Kerr-Newman configurations.  Innermost stable circular orbits
        (ISCOs) lie far away from the black hole horizon with respect
 to rotating black holes in GR.

 Having a new family of analytic rotating solutions for a theory of modified gravity opens new opportunities 
 for finding ways to test these theories against astrophysical or cosmological  observations. It will be interesting to determine
 whether there exists an analogue of the Carter constant \cite{Carter:1968rr}, which allows us to integrate the equations
 for time-like geodesics also beyond the case of circular equatorial orbits we studied here.  The analysis of possible
 mechanisms for extracting  rotational energy from the black hole, possibly exploiting vector interactions, deserves
  further
   study.  Finally, the issue of stability of our configurations is an open issue that  will need to be addressed
   for understanding the physical relevance of these objects. 
   
   
\subsubsection*{Acknowledgments} 
It is a pleasure to thank Javier Chagoya for discussions and suggestions. We are also grateful 
to  Tsvetan Vetsov for  pointing out some errors in an earlier 
 version of this paper.

\begin{appendix}

\section{
Asymptotic mass,  charge and angular momentum}\label{app-A}


For rotating configurations which admit  time-like Killing vectors, as is our case, we
can define conserved currents that can be used to obtain asymptotic charges: see for example \cite{Hartle:2003yu} for details.

The gravitational energy -- the mass -- can be obtained starting from 
 the time-like Killing vector $k^{\mu}$ associated with time translations, $k^{\mu}=\{1,0,0,0 \}$, by  defining a current 
 as
 follows
\beq
J^{\mu}=k_{\nu}R^{\mu\nu}
\eeq
Such current is conserved by means of  Killing condition $\nabla_{[\mu}k_{\nu]}=0$ and the Bianchi identity. We can use it to define
a conserved energy
\beq
E_{BH}= \int_{\Sigma}{d^3x\, \sqrt{\gamma^{(3)}}\,n_{\mu}J_{M}^{\mu}} \label{eq:ene}
\eeq 
where $\Sigma$ is a space-like hypersurface with induced metric $\gamma_{ij}^{(3)}$ and $n_{\mu}$ is the unit normal vector to $\Sigma$. Using Stokes theorem, and the properties of Killing vectors, we can rewrite this quantity as
\beq
E_{BH} = \int_{\partial \Sigma}{d^2x\, \sqrt{\gamma^{(2)}}\,n_{\mu}\,\sigma_{\nu}\nabla^{\mu}k^{\nu}},
\eeq
where the boundary $\partial \Sigma$ has metric $\gamma_{ij}^{(2)}$ and an outward-pointing normal vector $\sigma^{\mu}$.
 In our case, a computation of this quantity for our configuration \eqref{dis-sol1} gives $E_M \,= \, M$, hence the parameter $M$
 entering in the geometry corresponds to the gravitational mass of the object.
 
 Analogously, we can define a conserved `electric' charge of the object
\beq
Q_{BH}=-\int_{\partial \Sigma}{d^2x\, \sqrt{\gamma^{(2)}}\,n_{\mu}\,\sigma_{\nu}F^{\mu\nu}}
\eeq
which in our case gives, as expected,  $Q_{BH}\,=\,Q$, since at spatial infinity we do not expect contributions from the scalar field. In fact, the geometry does not feel the disformal transformation at spatial infinity, since the "disformal contribution" goes to infinity much faster than the GR one. \\ 
Finally, the conserved angular momentum is 
\be
J_{BH}\,=\,-\frac{1}{2} \int_{\partial \Sigma}{d^2x\, \sqrt{\gamma^{(2)}}\,n_{\mu}\,\sigma_{\nu}\nabla^{\mu}\varphi^{\nu}},
\eeq
where $\varphi^{\mu}$ is a  Killing vector associated with rotational symmetry around the coordinate $\phi$:
 $\varphi^{\mu}=\{ 0,0,0,1\}$.
 For our configuration we find $J_{BH}\,=\,a\,M$.
 
 To sum up, we find three  conserved asymptotic quantities with a clear physical interpretation, which can be easily associated
  with the constant parameters appearing in the geometry. 
  
  
\end{appendix}

\end{document}